\documentstyle[prl,aps,floats,epsf,times,latexsym]{revtex}
\newcommand{\makepreprint}{0}

\newcommand{\maketable}{1}

\def\np#1#2#3   {{ Nucl. Phys.} {\bf#1}, #2 (#3). }
\def\pcps#1#2#3 {{ Proc. Cam. Phil. Soc.} {\bf#1}, #2 (#3). }
\def\pl#1#2#3   {{ Phys. Lett.} {\bf#1}, #2 (#3). }
\def\plc#1#2#3   {{ Phys. Lett.} {\bf#1}, #2 (#3); }
\def\prep#1#2#3 {{ Phys. Rep.} {\bf#1}, #2 (#3). }
\def\prev#1#2#3 {{ Phys. Rev.} {\bf#1}, #2 (#3). }
\def\prl#1#2#3  {{ Phys. Rev. Lett.} {\bf#1}, #2 (#3). }
\def\prs#1#2#3  {{ Proc. Roy. Soc.} {\bf#1}, #2 (#3). }
\def\ptp#1#2#3  {{ Prog. Th. Phys.} {\bf#1}, #2 (#3). }
\def\rmp#1#2#3  {{ Rev. Mod. Phys.} {\bf#1}, #2 (#3). }
\def\rpp#1#2#3  {{ Rep. Prog. Phys.} {\bf#1}, #2 (#3). }
\def\zp#1#2#3   {{ Zeit. Phys.} {\bf#1}, #2 (#3). }
\def\epj#1#2#3   {{ Eur. Phys. Jour.} {\bf#1}, #2 (#3). }
\def\nim#1#2#3   {{ Nucl. Instr. Meth.} {\bf#1}, #2 (#3). }

\newcommand{\rmt}{\rm\textstyle}
\newcommand{\rms}{\rm\scriptstyle}
\newcommand{\stw}{\mbox{$\sin^2\theta_W$}}
\newcommand{\stwos}{\mbox{$\sin^2\theta_W^{\rms(on-shell)}$}}
\newcommand{\nub}{\overline{\nu}}

\newcommand{\qbar}[0]{\overline{q}}

\newcommand{\sbar}[0]{\overline{s}}

\newcommand{\uubar}[0]{\mbox{$\stackrel{(-)}{u}$}}
\newcommand{\ddbar}[0]{\mbox{$\stackrel{(-)}{d}$}}
\newcommand{\nubar}[0]{\overline{\nu}}
\newcommand{\numu}[0]{\nu_{\mu}}

\newcommand{\nube}[0]{\overline{\nu}_{e}}
\newcommand{\nue}[0]{\nu_{e}}

\newcommand{\nubmu}[0]{\overline{\nu}_{\mu}}

\newcommand{\nunub}{\stackrel{{\footnotesize (-)}}{\nu}}

\newcommand{\Kethree}{\mbox{$K^{\pm}_{e3}$}}
 
\newcommand{\Rnu}{\mbox{$R^{\nu}$}}
\newcommand{\Rnub}{\mbox{$R^{\nub}$}}

\newcommand{\Rmeasnu}{\mbox{$R_{\rms exp}^{\nu}$}}
\newcommand{\Rmeasnub}{\mbox{$R_{\rms exp}^{\nub}$}}
\newcommand{\ecal}{\mbox{$E_{\rms cal}$}}
\newcommand{\gLeff}{\mbox{$g_L^{\rms eff}$}}
\newcommand{\gReff}{\mbox{$g_R^{\rms eff}$}}
\newcommand{\gLReff}{\mbox{$g_{L,R}^{\rms eff}$}}
\newcommand{\uLReff}{\mbox{$u_{L,R}^{\rms eff}$}}
\newcommand{\dLReff}{\mbox{$d_{L,R}^{\rms eff}$}}

\begin{document}

\title{A Precise Determination of Electroweak Parameters 
in Neutrino-Nucleon Scattering}

\author{G.~P.~Zeller$^{5}$, K.~S.~McFarland$^{8,3}$,   
 T.~Adams$^{4}$, A.~Alton$^{4}$, S.~Avvakumov$^{8}$, 
 L.~de~Barbaro$^{5}$, P.~de~Barbaro$^{8}$, R.~H.~Bernstein$^{3}$, 
 A.~Bodek$^{8}$, T.~Bolton$^{4}$, J.~Brau$^{6}$, D.~Buchholz$^{5}$, 
 H.~Budd$^{8}$, L.~Bugel$^{3}$, J.~Conrad$^{2}$, R.~B.~Drucker$^{6}$, 
 B.~T.~Fleming$^{2}$, R.~Frey$^{6}$, J.A.~Formaggio$^{2}$, J.~Goldman$^{4}$, 
 M.~Goncharov$^{4}$, D.~A.~Harris$^{8}$, R.~A.~Johnson$^{1}$, J.~H.~Kim$^{2}$,
 S.~Koutsoliotas$^{2}$, M.~J.~Lamm$^{3}$, W.~Marsh$^{3}$, D.~Mason$^{6}$, 
 J.~McDonald$^{7}$, C.~McNulty$^{2}$, 
    D.~Naples$^{7}$, 
 P.~Nienaber$^{3}$, A.~Romosan$^{2}$, W.~K.~Sakumoto$^{8}$, H.~Schellman$^{5}$,
 M.~H.~Shaevitz$^{2}$, P.~Spentzouris$^{2}$, E.~G.~Stern$^{2}$, 
 N.~Suwonjandee$^{1}$, M.~Tzanov$^{7}$, M.~Vakili$^{1}$, A.~Vaitaitis$^{2}$, 
 U.~K.~Yang$^{8}$, J.~Yu$^{3}$, and E.~D.~Zimmerman$^{2}$}
\address{
$^1$University of Cincinnati, Cincinnati, OH 45221 \\
$^2$Columbia University, New York, NY 10027 \\
$^3$Fermi National Accelerator Laboratory, Batavia, IL 60510 \\
$^4$Kansas State University, Manhattan, KS 66506 \\
$^5$Northwestern University, Evanston, IL 60208 \\
$^6$University of Oregon, Eugene, OR 97403 \\
$^7$University of Pittsburgh, Pittsburgh, PA 15260 \\
$^8$University of Rochester, Rochester, NY 14627 \\ 
}
\date{\today}
\maketitle
\begin{abstract}

The NuTeV collaboration has extracted the electroweak parameter $\stw$ from 
the measurement of the ratios of neutral current to charged current $\nu$ and 
$\nub$ cross-sections.  Our value, 
$\stw^{({\rms on-shell)}}=0.2277\pm0.0013({\rmt stat})\pm0.0009({\rmt syst})$,
is 3 standard deviations above the standard model prediction.  
We also present a model independent analysis of the same data 
in terms of neutral-current quark couplings.

\end{abstract} 
\ifnum\makepreprint=0
\twocolumn
\fi

%

Neutrino-nucleon scattering is one of the most precise probes of the weak
neutral current.  The Lagrangian for weak neutral current $\nu$--$q$
scattering can be written
as
\begin{eqnarray}
{\cal L}&=&-\frac{G_F\rho_0}{\sqrt{2}}(\nubar\gamma^\mu(1-\gamma^5)\nu)
        \nonumber\\
          &&\times\left( \epsilon^q_L {\qbar}\gamma_\mu(1-\gamma^5){q}+
                \epsilon^q_R {\qbar}\gamma_\mu(1+\gamma^5){q}\right) , 
\label{eqn:lagrangian}
\end{eqnarray}
where deviations from $\rho_0=1$ describe non-standard sources of
SU(2) breaking, and $\epsilon^q_{L,R}$ are the chiral quark
couplings.  For the weak charged current, $\epsilon^q_L=I_{\rms weak}^{(3)}$
and $\epsilon^q_R=0$, but for the neutral current $\epsilon^q_L$ and
$\epsilon^q_R$ each contain an additional term, $-Q\stw$, where $Q$ is the 
quark's electric charge in units of $e$.  By measuring ratios of the charged 
and neutral current processes on a hadronic target, one can thus extract $\stw$
and $\rho_0$.


In the context of the standard model, this measurement of $\stw$
is comparable in precision to direct measurements of $M_W$.  Outside of the 
standard model, neutrino-nucleon scattering provides one of the most precise 
constraints on the weak couplings of light quarks, and tests the validity of 
electroweak theory in a range of momentum transfer far from $M_Z$.  This process is also 
sensitive to non-standard interactions, including possible contributions from
leptoquark and  $Z^\prime$ exchange\cite{langacker
}.

The ratio of
neutral current to charged current cross-sections 
for either $\nu$ or $\nub$ scattering
from isoscalar targets of $u$ and $d$ quarks can be written as\cite{llewellyn}
\begin{equation}
R^{\nu(\nub)} \equiv \frac{\sigma(\nunub N\rightarrow\nunub X)}
                 {\sigma(\nunub N\rightarrow\ell^{-(+)}X)}  
= (g_L^2+r^{(-1)}g_R^2),
\label{eqn:ls}
\end{equation}
where
\begin{equation}
r \equiv \frac{\sigma({\overline \nu}N\rightarrow\ell^+X)}
                {\sigma(\nu N\rightarrow\ell^-X)} \sim \frac{1}{2},  
\label{eqn:rdef} 
\end{equation}
and $g_{L,R}^2=(\epsilon^u_{L,R})^2+(\epsilon^d_{L,R})^2$.
Corrections to Equation~\ref{eqn:ls} result from the presence of heavy quarks 
in the sea, the production of heavy quarks in the target, higher order 
terms in the cross-section, and any isovector component of the light quarks 
in the target.  In particular, in the case where a final-state charm quark is 
produced from a $d$ or $s$ quark in the nucleon, there are large uncertainties 
resulting from the mass suppression of the charm quark.  This uncertainty has
limited the precision of previous measurements of electroweak parameters in
neutrino-nucleon scattering\cite{CCFR,CDHS,CHARM}.

To reduce the effect of uncertainties resulting from charm production, Paschos
and Wolfenstein\cite{Paschos-Wolfenstein} suggested consideration of the
observable:
 
\begin{eqnarray}
R^{-} &\equiv& \frac{\sigma(\nu_{\mu}N\rightarrow\nu_{\mu}X)-
                   \sigma(\nub_{\mu}N\rightarrow\nub_{\mu}X)}
                  {\sigma(\nu_{\mu}N\rightarrow\mu^-X)-  
                   \sigma(\nub_{\mu}N\rightarrow\mu^+X)} \nonumber\\  
&=& \frac{\Rnu-r\Rnub}{1-r}=(g_L^2-g_R^2).
\label{eqn:rminus}
\end{eqnarray}

\noindent
$R^-$ is more difficult to measure than $R^\nu$, primarily because the neutral
current scatterings of $\nu$ and $\nub$ yield identical observed final
states which can only be distinguished through {\em a priori} knowledge of the
initial state neutrino.

\section*{Method}
High-purity $\nu$ and $\nub$ beams were provided by the Sign Selected
Quadrupole Train (SSQT) beamline at the Fermilab Tevatron during the 1996-1997
fixed target run. Neutrinos were produced from the decay of pions and kaons
resulting from interactions of $800$~GeV protons in a BeO target. Dipole
magnets immediately downstream of the proton target bent pions and kaons of
specified charge in the direction of the NuTeV detector, while oppositely
charged and neutral mesons were stopped in beam dumps. The resulting beam was
almost pure $\nu$ or $\nub$, depending on the charge of the parent
mesons. Anti-neutrino interactions comprised $0.03$\% of the neutrino 
beam events, and neutrino interactions $0.4$\% of the anti-neutrino 
beam events. In addition, the beams of almost pure muon neutrinos contained a 
small component of electron neutrinos (mostly from $\Kethree$ decays) which 
created $1.7\%$ of the observed interactions in the neutrino beam and $1.6\%$ 
in the anti-neutrino beam.

Neutrino interactions were observed in the NuTeV detector\cite{nim}, located
1450 m downstream of the proton target. The detector consisted
of an 18 m long, 690 ton steel-scintillator target, followed by an 
iron-toroid spectrometer. The target calorimeter was composed of 168 (3m 
$\times$ 3m $\times$ 5.1cm) steel plates interspersed with liquid 
scintillation counters (spaced every two plates) and drift chambers (spaced 
every four plates). The scintillation counters provided triggering information 
as well as a measurement of the longitudinal interaction vertex, event length, 
and energy deposition. The mean position of hits in the drift chambers 
established the transverse vertex for the event. The toroid spectrometer,
used to determine muon charge and momentum, also provided a measurement of
the muon neutrino flux in charged current events. In addition, the detector 
was calibrated continuously through exposure to beams of hadrons, electrons, 
and muons over a wide energy range~\cite{nim}.

For inclusion in this analysis, events are required to deposit at least 20
GeV of visible energy ($\ecal$) in the calorimeter, which ensures full
efficiency of the trigger, allows an accurate vertex determination, and 
reduces cosmic ray background. Events with $\ecal>180$ GeV are also
removed. Fiducial criteria restrict the location of the neutrino interaction 
to the central region of the calorimeter. The chosen 
fiducial volume enhances interactions that are contained in the calorimeter, 
and minimizes the fraction of events from electron neutrinos or non-neutrino 
sources. After all selections, the resulting data sample consists of 
1.62 $\times$ 10$^{6}$ $\nu$ and 0.35 $\times$ 10$^{6}$ $\nub$ events with 
a mean visible energy ($\ecal$) of 64 GeV and 53 GeV, respectively.



In order to extract $\stw$, the observed neutrino events must be separated
into charged current (CC) and neutral current (NC) candidates. Both CC and
NC neutrino interactions initiate a cascade of hadrons in the target that
is registered in both the scintillation counters and drift chambers. 
Muon neutrino CC events are distinguished by the presence of a final
state muon that typically penetrates beyond the hadronic shower
and deposits energy in a large number of consecutive scintillation counters.
NC events usually have no final state muon and deposit energy over a
range of counters typical of a hadronic shower.

These differing event topologies enable the statistical separation of
CC and NC neutrino interactions based solely on event length. For each event, 
this length is defined by the number of scintillation counters between the
interaction vertex and the last counter consistent with at least single muon 
energy deposition. Events with a ``long'' length are identified as CC 
candidates, while ``short'' events are most likely NC induced.  The separation
between short and long events is made at 16 counters ($\sim$ 1.7m of steel) 
for $\ecal \leq 60$ GeV, at 17 counters for $60 < \ecal \leq 100$ GeV, and 
otherwise at 18 counters. The ratios of short to long events measured in 
the $\nu$ and $\nub$ beams are:

\begin{equation}
\Rmeasnu= 0.3916 \pm 0.0007~{\rm and}~\Rmeasnub= 0.4050 \pm 0.0016.
\end{equation}

\noindent
$\stw$ can be extracted directly from these measured ratios 
by comparison with a detailed Monte Carlo simulation of the experiment. 
The Monte Carlo must include neutrino fluxes, the neutrino cross-sections, 
and a detailed description of the detector response. 

A detailed beam simulation is used to predict the $\nu$ and $\nub$ 
fluxes. In particular, a precise determination of the electron neutrino
contamination in the beam is essential. The ratios $\Rmeasnu$ and
$\Rmeasnub$ increase in the presence of electron neutrinos 
in the data sample because electron neutrino charged current interactions are almost 
always identified as neutral current interactions. 

The bulk of the observed electron neutrinos, 93$\%$ in the $\nu$ beam and 
70$\%$ in the $\nub$ beam, result from $\Kethree$ decays.  The beam
simulation can be tuned with high accuracy to describe $\nu_{e}$ and 
$\nub_{e}$ production from charged kaon decay because the $K^{\pm}$ 
contribution is constrained by the observed $\nu_\mu$ and $\nub_\mu$ fluxes. 
Because of the precise alignment of the beamline elements and the low 
acceptance for neutral particles, the largest uncertainty in the calculated 
electron neutrino flux is the 1.4\% uncertainty in the $\Kethree$ branching 
ratio\cite{pdg}.  Other sources of electron neutrinos include neutral kaons, 
charmed hadrons, and muon decays, all of which have larger fractional 
uncertainties ($10$--$20$\%).  Finally, small uncertainties in the calibration
of the calorimeter and the muon toroid affect the muon and electron neutrino 
flux measurements.  Additional constraints from the data, including direct 
measurements of $\nue$ and $\nube$ charged current events and measurements of 
$\numu$ events in the $\nubmu$ beam (which also result from charm and neutral 
kaon decay)\cite{drew} reduce the electron neutrino uncertainties. At the 
highest energies (E$_\nu>$350 for $\numu$ and E$_\nu>$180 for $\nue$), the 
beam Monte Carlo underpredicts the measured flux and is thus not used.

Neutrino-nucleon deep inelastic scattering processes are simulated using a
leading order (LO) model for the cross-section augmented with longitudinal 
scattering and higher twist terms. The cross-section parameterization 
incorporates LO parton distribution functions (PDFs) from charged current
data measured obtained with the same target and model as used in this 
experiment\cite{buras,ukcomb2}.  These PDFs include an external constraint on
$\sigma^{\nub}/\sigma^{\nu}$\cite{ukcomb2}, 
and make the standard assumptions that 
$\uubar_p(x)=\ddbar_n(x)$, $\ddbar_p(x)=\uubar_n(x)$ and $s(x)=\sbar(x)$.  
Small modifications adjust the
parton densities to produce the inherent up-down quark asymmetry consistent
with muon scattering\cite{nmc} and Drell-Yan\cite{e866} data. A LO analysis
of $\nunub N\to\mu^+\mu^-X$ events\cite{max} provides the
shape and magnitude of the strange sea.  Mass suppression from charged
current charm production is modeled using a LO slow rescaling 
formalism\cite{slowrescaling} whose parameters and uncertainties come from 
the same high-statistics $\mu^+\mu^-$ sample. A model for $c\overline{c}$ 
production is chosen to match EMC data\cite{emc}; it is assigned a 100$\%$ 
uncertainty. A global analysis\cite{whitlow} provides a parameterization of 
the longitudinal structure function, R$_{L}$, which is allowed to vary within 
its experimental and theoretical uncertainties. QED and electroweak radiative
corrections to the scattering cross-section are applied using code supplied 
by Bardin\cite{bardin} and from V6.34 of ZFITTER\cite{zfitter}, 
and uncertainties are estimated by varying the 
parameters in these corrections.

The Monte Carlo must also accurately simulate the response of the detector
to the products of neutrino interactions in the target. The critical
parameters that must be modeled are the calorimeter response to muons, the
measurement of the position of the neutrino interactions, and the
range of hadronic showers in the calorimeter. Precise determination
of these effects is made through extensive use of both neutrino and 
calibration beam data. Measured detector parameters are then varied within 
their uncertainties to estimate systematic errors.

\begin{figure}[tbp]
\centerline{\epsfxsize 3.5 truein \epsfbox[3 8 555 551]{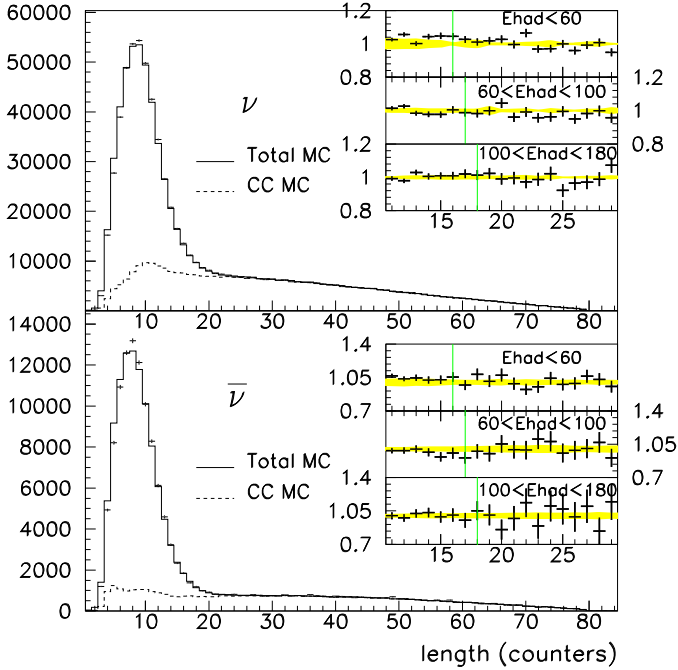}}   
 \caption{Comparison of $\nu$ and $\nubar$ event length distributions in 
          data and Monte Carlo (MC). 
          The MC prediction for CC events is shown separately. 
          Insets show data/MC ratio comparisons in the region of the length 
          cut with bands to indicate the $1\sigma$ systematic uncertainty
          in this ratio.}
 \label{fig:length}
\end{figure}

An important test of the simulation is its ability to predict the length
distribution of events. Figure~\ref{fig:length} shows event length 
distributions in the final data sample compared to the Monte Carlo
prediction for our measured value of $\stw$. Events reaching the toroid, which
comprise about 80$\%$ of the CC sample, have been left out for clarity, but
are included in the normalization of the data. Excellent agreement within
uncertainties is observed in the overlap region of long NC and short CC events.

\section*{Results}

Having precisely determined $\Rmeasnu$, $\Rmeasnub$, and their predicted values
as a function of electroweak parameters $\stw$ and $\rho_0$, we proceed to
extract the best values of $\stw$ and $\rho_0$. This is done by means of a fit 
that also includes the slow-rescaling mass for charm production ($m_c$) with 
its {\em a priori} constraint from $\mu^+\mu^-$ data\cite{max}.  $\Rnub$ is 
much less sensitive to $\stw$ than $\Rnu$, but both are sensitive to $m_c$ 
and $\rho_0$.

\begin{table}
\ifnum\maketable=1
 \begin{tabular}{|r|c||c|c|} 
 SOURCE OF UNCERTAINTY & $\delta\stw$ & $\delta\Rnu$ & $\delta\Rnub$ \\ \hline
 {Data Statistics}                      & {0.00135} & {0.00069} & {0.00159}   \\ 
 {Monte Carlo Statistics}               & {0.00010} & {0.00006} & {0.00010} \\ \hline
 { \bf TOTAL STATISTICS \hfill }        & {0.00135} & {0.00069} & {0.00159} \\ \hline\hline
%
 {$\nu_e, \nub_e$ Flux}                 & {0.00039} & {0.00025} & {0.00044} \\
 {Energy Measurement}                   & {0.00018} & {0.00015} & {0.00024} \\
 {Shower Length Model}                  & {0.00027} & {0.00021} & {0.00020} \\
 {Counter Efficiency, Noise, Size}      & {0.00023} & {0.00014} & {0.00006} \\
 {Interaction Vertex}                   & {0.00030} & {0.00022} & {0.00017} \\ \hline
 { \bf TOTAL EXPERIMENTAL\hfill }       & {0.00063} & {0.00044} & {0.00057} \\ \hline\hline
 {Charm Production, Strange Sea}        & {0.00047} & {0.00089} & {0.00184} \\
 {Charm Sea}                            & {0.00010} & {0.00005} & {0.00004} \\
 {$\sigma^{\nub}/\sigma^{\nu}$}         & {0.00022} & {0.00007} & {0.00026} \\
 {Radiative Corrections}                & {0.00011} & {0.00005} & {0.00006} \\
 {Non-Isoscalar Target}                 & {0.00005} & {0.00004} & {0.00004} \\
 {Higher Twist}                         & {0.00014} & {0.00012} & {0.00013} \\ 
 {$R_L$}                                & {0.00032} & {0.00045} & {0.00101} \\ \hline 
 { \bf TOTAL MODEL \hfill }             & {0.00064} & {0.00101} & {0.00212} \\ \hline\hline\hline
 { \bf TOTAL UNCERTAINTY \hfill }       & {0.00162} & {0.00130} & {0.00272} 
 \end{tabular}
\fi
\caption{Uncertainties for both the single parameter $\stw$ 
        fit and for the comparison of $\Rnu$ and $\Rnub$ with model 
        predictions.}
 \label{table:systematics}
\end{table}

When fitting with the assumption $\rho_0=1$, $\stw$ is
simultaneously fit with the slow-rescaling parameter $m_c$.  Like an explicit
calculation of $R^-$, this procedure reduces uncertainties related to sea 
quark scattering as well as many experimental systematics common to both $\nu$ 
and $\nub$ samples.  Statistical and systematic uncertainties in the 
$\stw$ fit and in the comparison of $\Rnu$ and $\Rnub$ with the Monte Carlo 
prediction are shown in Table~\ref{table:systematics}.

The single parameter fit for $\stw$ measures:

\begin{eqnarray}
    \sin^2\theta_W^{({\rms on-shell)}}&=&0.22773\pm0.00135({\rmt stat.})\pm0.00093({\rmt syst.})
        \nonumber\\
        &-&0.00022\times(\frac{M_{top}^2-(175 \: \mathrm{GeV})^2}{(50 \: \mathrm{GeV})^2})
        \nonumber\\
        &+&0.00032\times \ln(\frac{M_{Higgs}}{150 \: \mathrm{GeV}})
\end{eqnarray}

\noindent
Leading terms in the one-loop electroweak radiative corrections\cite{bardin}
produce the small residual dependence of our result on M$_{top}$ and 
M$_{Higgs}$.  The prediction from 
the standard model with parameters determined by 
a fit to other electroweak 
measurements is 
$0.2227\pm0.0004$\cite{LEPEWWG,Martin}, approximately $3\sigma$ from our 
result. In the on-shell scheme, where 
$\sin^2\theta_W\equiv 1-M_{W}^2/M_{Z}^2$, and where $M_{W}$ and $M_{Z}$ are 
the physical gauge boson masses, our result implies $M_{W}=80.14\pm0.08$~GeV. 
The world-average of the direct measurements of $M_W$ is 
$80.45\pm0.04$~GeV\cite{LEPEWWG}. 


For the simultaneous fit to $\stw$ and $\rho_0$, we obtain:
\begin{equation}
\rho_0=0.99789\pm 0.00405,~\stw=0.22647\pm 0.00311,
\end{equation}
with a correlation coefficient of $0.850$ between the two parameters.  This
suggests one but not both of $\stwos$ or $\rho_0$ may be consistent with
expectations. We have also performed a two-parameter fit in 
terms of the isoscalar combinations\footnote{Due to the asymmetry between the 
strange and charm seas and to the slight excess of neutrons in our target,
this result is only sensitive to isovector combinations at about $3\%$ of 
the sensitivity of isoscalar couplings.} of effective\footnote{Effective
couplings are those which describe observed experimental rates when the
processes described by Eqn.~\ref{eqn:lagrangian} are calculated without
electroweak radiative corrections.}  neutral-current quark 
couplings $(\gLReff)^2=(\uLReff)^2+(\dLReff)^2$ at 
$\left< q^2\right> \approx-20$~GeV$^2$, which yields:
\begin{equation}
(\gLeff)^2=0.30005\pm0.00137,~(\gReff)^2=0.03076\pm0.00110,
\end{equation}
with a correlation coefficient of $-0.017$. 
The predicted values from Standard Model parameters corresponding to the
electroweak fit described earlier\cite{LEPEWWG,Martin}
are
$(\gLeff)^2=0.3042$ and
$(\gReff)^2=0.0301$.

In conclusion, NuTeV has made precise determinations of the electroweak
parameters through separate measurements of 
$\Rnu$ and $\Rnub$.  We find a significant disagreement with the
standard model expectation for $\stwos$.  In a model-independent analysis,
this result suggests a smaller left-handed neutral current coupling to the 
light quarks than expected.


\section*{Acknowledgements}
We thank the staff of the Fermilab Beams, Computing and Particle 
Physics Divisions for design, construction, and operational assistance during 
the NuTeV experiment.  This work was supported by the U.S. Department of 
Energy, the National Science Foundation and the Alfred P. Sloan foundation.

%

\end{document}